# Exceptional Lie Groups, E-infinity Theory and Higgs Boson

Ayman A. El-Okaby[*]

Department of Physics, Faculty of Science, Alexandria University, Egypt.

*elokaby@yahoo.com*

**Abstract**

In this paper we study the correlation between El-Naschie's exceptional Lie groups hierarchies and his transfinite E-Infinity spacetime theory. Subsequently this correlation is used to calculate the number of elementary particles in the standard model, mass of the Higgs bosons and some coupling constants.

## 1. Introduction

The standard model of elementary particles (SM) has passed every experimental challenge it met [1]. The most important aspect of the standard model which has not yet been verified experimentally is the Higgs sector. The existence of Higgs boson is supposed to share in solving some remaining mysteries of nature, particularly, those dealing with the origin of mass and how the elementary particles acquire it [1-4].

In the last few years, many theories dealt with the Higgs boson problem [5-8]. However in our opinion, one of the most promising ways to tackle this problem is E-infinity theory developed by the Egyptian physicist Mohamed Saladdin EL-Naschie [9-12]. It is the most recent theory that postulates infinitely many hierarchical extra-dimensions for quantum spacetime. In E-infinity theory space–time is assumed to be an infinite dimensional Cantor set. The bijection formula of E-infinity theory is given in [13] as:

$$d_c^{(n)} = (1/d_c^{(0)})^{n-1},$$

where $d_c^{(n)}$ is the Hausdorff dimension in n-dimensions, and $d_c^{(0)}$ is the core Menger-Uhyson zero dimensional set. Setting n=4 and $d_c^{(0)}$ equal to golden mean $\phi$ = 0.618033989 [11], one finds $d_c^{(4)} = (1/\phi)^3 = 4+\phi^3 = 4.236067977$. This value plays a fundamental role in E-infinity theory and is considered to be the expectation value of the Hausdorff dimension of E-infinity spacetime. However, the topological dimension associated with $4+\phi^3$ is exactly 4 and that although the formal dimension is infinity $n_f = \infty$ [9]. Thus E-infinity space-time is defined not only by one but also by three dimensions.

In the following, we will show that E-infinity theory in conjunction with holographic principle and El-Naschie's exceptional Lie groups hierarchies $E_8$, $E_7$ and $E_6$ provides an answer for some remaining mysteries in high energy physics, especially the number of the Higgs bosons and their masses [14-16].

## 2. Holographic principle and E-infinity theory

In 1993, Noble laureate Gerard 'tHooft proposed the dimensional reduction in quantum gravity theories. This is known now as the Holographic principle which was extended by Leonard Susskind. [17].

The Holographic principle is about encoding information in (D+1) dimensional space onto D-dimensional space. In other words, all the information contained in a volume of space can be represented by a theory that lives on the boundary of that region [17].

Within E-infinity theory, an extension to the transfinite equivalence is achieved by direct comparison and the addition or subtraction of small transfinite quantities. This is related to the golden mean $\phi = 0.618033989$ such as k= $\phi^3$ (1- $\phi^3$) = 0.18033989 and $k_o$ = $\phi^5$(1- $\phi^5$) = 0.082039325 that are used to extend $\bar{\alpha}_o$ =137 to $\bar{\alpha}_o$ = 137. 082039325 and $\bar{\alpha}_{gs}$ = 26 to $\bar{\alpha}_{gs}$ =26.18033989 [9]. The theory was extended to quantum golden field theory by Mohamed EL-Naschie and one of his students Nasr Ahmed now at New Castle University in UK [18, 19]

This idea can be applied to Klein modular curve $\Gamma(7)$ which could be seen as a topological deformation of $E_8$ exceptional symmetry group. The original curve has 336-fold symmetries corresponding exactly to 336 triangular pieces of which it is made. These 336 triangles are considered to be degree of freedom or dimension. Thus one can write the original Klein modular curve [20] as:

$$N(\text{symmetries}) = \text{Dim } \Gamma(7) = 2 (168) = 336.$$

This value is exactly equal to the number of independent components of the Riemannian tensor in 8-dimemsional super-space $R^{(8)}$ given by the familiar expression [13]:

$$R^{(8)} = \frac{n^2(n^2-1)}{12} = \frac{8^2(8^2-1)}{12} = 336$$

If $\Gamma(7)$ is compactified by adding infinite numbers of hyperbolic progressively smaller triangles (see Fig. 1), the formal dimension becomes infinite, but the expected transfinite value can be calculated according to E-infinity theory [21] as follows,

$$\text{Dim } \Gamma_c(7) = 336 + 16k = 338.8854382.$$

Each triangle in $\Gamma_c(7)$ is exactly of the same size, but because they are scaled down. They appear infinitely smaller as we tend towards the circular boundary, which we cannot ever reach. In several occasions E-infinity theory showed that the information contained in the 2- dimension $\Gamma_c(7)$ is contained in the full dimensional E-infinity. Thus one can say that $\Gamma_c(7)$ is assumed to be the holographic projection of E-infinity theory. [22].

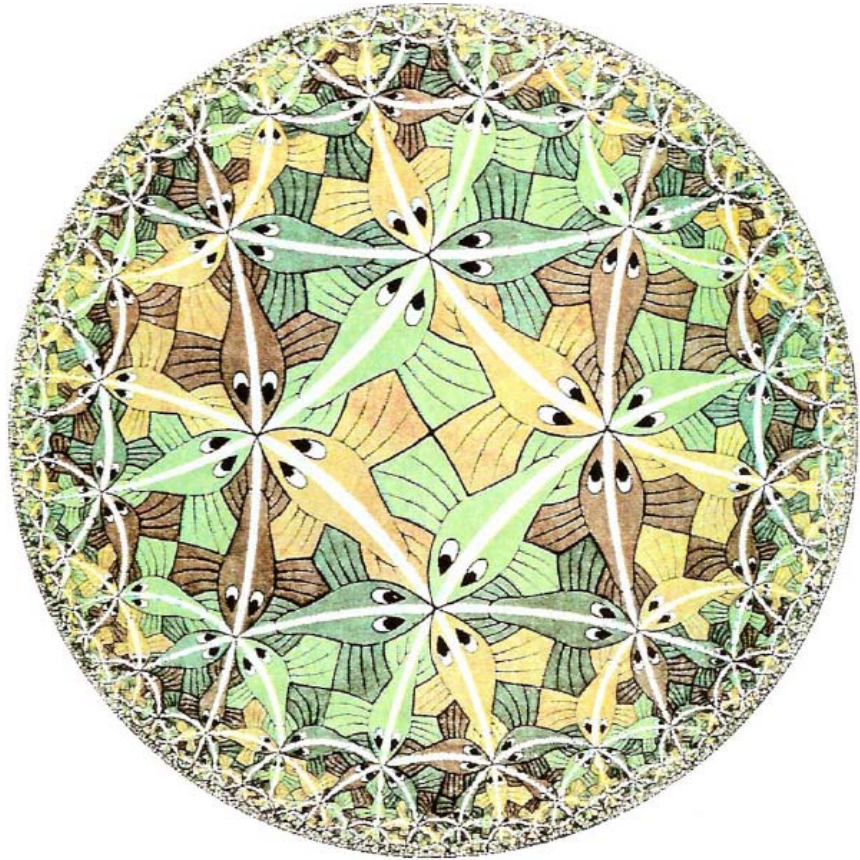

*Fig. (1) .Dutch artist M.C. Escher was introduced to the concept of hyperbolic tiling by the great geometer Donald Coxeter. With his "Circle Limit" series of drawings, Escher explored the infinite symmetries inherent in hyperbolic space by equal size fish tessellate their world in a symphony of triangles and squares. This curve has 336-fold symmetries correspond exactly to 336 triangular pieces of which it is made, if this curve is compactified by adding infinite numbers of hyperbolic progressively smaller triangles, the formal dimension becomes infinite, but the expected transfinite value will be Dim $\Gamma_c(7) = 336 + 16k \approx 339$. The similarity with Klein's $\Gamma_c(7)$ modular curve is obvious. This in turn is effectively a holographic boundary for EL-Naschie's Cantorian $\varepsilon^{(\infty)}$ theory space time [22]. Note, the topological similarity to El-Naschie's exceptional Lie groups hierarchy mapping of $E_8$ shown in [23].*

## 3. MSSM and the number of the Higgs particles

A minimal supersymmetric extension of the standard model (MSSM) by El-Naschie, predicted that the total number of the particles in the standard model is 66 particles. These are the known 60 experimentally confirmed particles, 5 massive spin zero Higgs bosons and one massless spin 2 graviton [11]. The same results could be obtained following El-Naschie's by considering a three steps symmetry breaking of $N_k^{(32)} = n(n+1)/2 = 528$, which is the number of killing's vector fields in the $n = (4)(8) = 32$ super-space of E-infinity theory [24], Thus:

$$N(SM) = 528/8 = 66 \text{ particles.}$$

EL-Naschie and other authors have showed, using a Fuzzy $K_3$ manifold and the theories of P-Brane and instantons [21], that a consistent super-symmetric extension of the standard model which includes gravity is likely to have a total number of elementary particles equal 66 particles [21]. However the maximum number of particles- like degree of freedom is equal to 69. That is what Prof. Ji-Huan He at Donghua University in Shanghai called the missing 9 elementary particles [16]

This maximum number is easily obtained using the exact transfinite theory of El-Naschie [25]:

$$N = \frac{(26+k)(84+4k)}{(32+2k)} = \frac{\bar{\alpha}_o}{2} \cong 69$$

where $\bar{\alpha}_o$ is the inverse of the fine structure constant, (84+4k) is the field strength, (32+2k) is the compactified spin and (26+k) = $\chi$ is the compactified Euler characteristic of the $K(\varepsilon^{(\infty)})$ fuzzy Kahler manifold of E-infinity and $k = \phi^3(1-\phi^3) = 0.18033989$ [19].

## 4. The role of symmetry and group theory

Symmetry is one of the most fundamental properties of nature. The branch of mathematics dealing with symmetry is the group theory. These groups are extremely important and play a fundamental role in particle physics [26].

The present paper will follow the pioneering work of Green, Schwarz, Witten and El-Naschie [23] on the exceptional lie groups $E_8$, $E_7$ and $E_6$ due to their important applications in wild topology and geometry that is related to the space-time of E-infinity theory, as well as the relation to Klein modular spaces and superstring theory.

## 4.1. Root system and representation of exceptional lie groups

Wilhelm Killing classified root systems in the 1890s [27]. He found 4 infinite classes of Lie algebras, labeled $A_n$, $B_n$, $C_n$, and $D_n$, where n=1,2,3.... He also found 5 more exceptional ones: $G_2$, $F_4$, $E_6$, $E_7$, and $E_8$.

*E₈ representation*

E$_8$ is an example of Lie groups. These groups were invented by the 19th-century Norwegian mathematician, Sophus Lie, to study symmetry underlying any symmetrical object, such as a sphere [28].

The E$_8$ root system consists of all vectors (called roots) ($a_1,a_2,a_3,a_4,a_5,a_6,a_7,a_8$) where all $a_i$ are integers or all integers plus 1/2. The sum is an even integer, sum of the squares is 2, and there are 240 of them, 112 roots corresponding to integers and 128 roots corresponding to half-integers [27].

Another way to describe the geometry of the 8-dimensional space is by extended Coxeter graph [28]

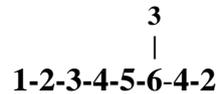

This graph represents 9 in balance vectors, all are in the 8-dimensional reflection space of E$_8$. The balance numbers are the lengths of these vectors, which are set at 120 degrees to each other whenever a line in the graph connects two balance numbers.

The sum of the E$_8$ balance numbers is 30, which is called the E$_8$ Coxeter number. If we multiply 30 by 8, we get 240, which is the numbers of roots assumed by E$_8$ root system [29].

Thus, the $E_8$ root system consists of 240 vectors in an eight-dimensional space. These vectors are the vertices (corners) of an eight-dimensional object called the Gosset polytope $4_{21}$ (Fig. 2) as assumed for the first time by John H. Conway [29].

These 240 vectors form the basis for the 240 non-commutative dimensions of the $E_8$ Lie algebra. There are more 8 commutative dimensions that make the total dimensions of $E_8$ Lie algebra is equal to 248. El-Naschie in his pioneering work on exceptional Lie groups hierarchy was able to interpret all these points and the number of fermions and bosons

These 240 vectors are considered to be the kissing points in the reflection space of $E_8$, which is the number of the 7- dimensional spheres that pack around a central sphere in the 8 –dimensional space. The 240 spheres touch that central sphere in 240 kissing points. Thus, one can say that 240 is the $E_8$ kissing number and this will be denoted following El-Naschie as $K(E_8)$ [23].

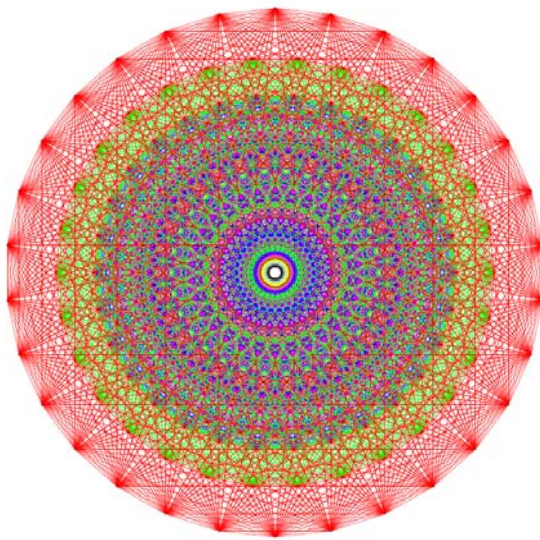 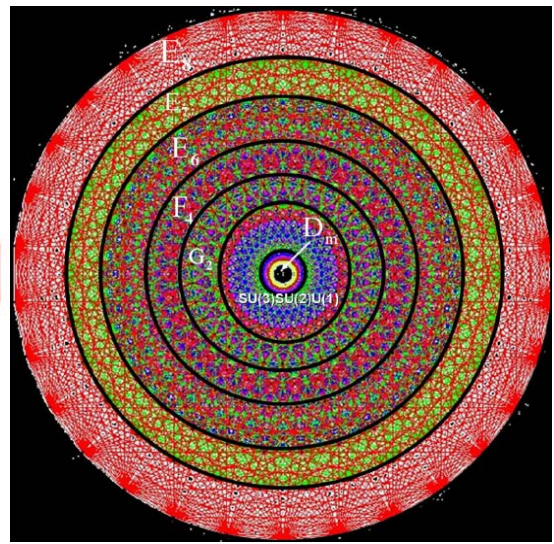

Fig. 2-a  Fig. 2-b

*Fig. (2-a) The image of Gosset polytope $4_{21}$ as generated by John Stembridge, based on McMullen's drawing. The lines in the picture connect adjacent vertices in the polytope, with colors chosen according to the length of the 2-dimensional projection. Since the picture is a 2-dimensional projection of an 8-dimensional object. Fig. (2-b) EL-Naschie view for an exceptional lie group hierarchy and the standard model in 11 dimensions projected on a two dimensional representation of $E_8$. Notes that this object is 57 dimensions and may be related to Klein modular curve and the fabric of the cosmos [23]. El-Naschie, extending the pioneering work of M. Green in Cambridge and J. Schwarz in CalTech, related these numbers to the massless gauge bosons as well as the elementary particles of the standard model [25].*

## $E_7$ representation

$E_7$ Lie group has dimensions of 133. That number is easily calculated by writing down the extended $E_7$ Coxeter graph [28]:

```
              2
              |
     1--2--3--4--3--2--1
```

The sum of the balance numbers is 18. If we multiply 18 by 7, we get 126 non-commutative dimensions these are the $E_7$ kissing numbers $K(E_7)$; now add the remaining 7 commutative dimensions. One finds 7 + 126 = 133 dimensions.

## $E_6$ representation

Following the same procedure mentioned earlier one can easily calculate the $E_6$ kissing number $K(E_6)$ to be equal 72. Thus, the total dimensionality of the $E_6$ Lie group is 78.

The dimensions, ranks, roots and kissing numbers of different lie groups are summarized in Table 1 and Table 2.

*Table 1. Exceptional Lie groups.*

| Lie group | Dimension | Rank |
|---|---|---|
| $G_2$ | 14 | 2 |
| $F_4$ | 52 | 4 |
| $E_6$ | 78 | 6 |
| $E_7$ | 133 | 7 |
| $E_8$ | 248 | 8 |
| SUM | 525 | 27 |

*Table 2. A-D-E Series of Lie groups.*

| Lie group | Dimension | Sum of balance numbers | Number of roots | Kissing numbers |
|---|---|---|---|---|
| $A_0$ | 1 | 1 | 1 | 1 |
| $A_1$ | 3 | 2 | 2 | 2 |
| $A_2$ | 8 | 3 | 6 | 6 |
| $D_3$ | 15 | 4 | 12 | 12 |
| $D_4$ | 28 | 6 | 24 | At least 24 at most 25 |
| $D_5$ | 45 | 8 | 40 | at least 40 at most 46 |
| $E_6$ | 78 | 12 | 72 | at least 72 at most 82 |
| $E_7$ | 133 | 18 | 126 | at least 126 at most 140 |
| $E_8$ | 248 | 30 | 240 | 240 |
| SUM | 559 | 84 | 523 | 523 |

*Note that, in our calculations we use the least values of the kissing numbers of the exceptional lie groups.*

## 5. Exceptional Lie groups hierarchy and SM

Based on Table 1 and Table 2, one can deduce the following A-D-E exceptional Lie group hierarchy [30, 31],

$$E_8, \ E_7, \ E_6, \ D_5, \ D_4, \ D_3, \ A_2, \ A_1, \text{ and } A_0$$

First, let us consider the sum of this hierarchy

$$\sum_{i=0}^{2} A_i + \sum_{i=3}^{5} D_i + \sum_{i=6}^{8} E_i = 12 + 88 + 459 = 559.$$

This sum of Lie groups hierarchy possesses highly interesting properties related to the standard model of elementary particles (SM) and that will be explained following El-Naschie's methodology [32].

The first term in the sum can be interpreted as follows,

$$\sum_{i=0}^{2} A_i = A_0 + A_1 + A_2 = 1 + 3 + 8 = 12.$$

Consequently, we have 12 states corresponding to 12 gauge bosons that are described in the standard model by gauge group SU(3) × SU(2) × U(1) , namely,

$$A_0 + A_1 + A_2 = U(1) + SU(2) + SU(3).$$

More specifically,

$A_0 = U(1) = 1 \Rightarrow$ Photon,

$A_1 = SU(2) = 2^2 - 1 = 3 \Rightarrow W^\pm \text{ and } Z^0$, and

$A_2 = SU(3) = 3^2 - 1 = 8 \Rightarrow$ Gluons.

Next, let us try to interpret the second term of the hierarchy namely,

$$\sum_{i=3}^{5} D_i = D_3 + D_4 + D_5 = 15 + 28 + 45 = 88.$$

We notice that,

$D_3 + D_5 = 60 \Rightarrow$ corresponds to 60 experimentally confirmed particles of the standard model.

$D_4 = 28 = 12 + 16$,

where,

$12 \Rightarrow SU(3) \times SU(2) \times U(1) = |SM|$,

and

$16 \Rightarrow$ missing Higgs boson + Graviton.

In the same time, we note that

$$(D_3 + D_5) - \sum_{i=0}^{2} A_i = 60 - 12 = 48$$

which is exactly equal to the number of fermions in the standard model.

Finally, let us move to the third term in the sum,

$$\sum_{i=6}^{8} E_i = E_6 + E_7 + E_8 = 78 + 133 + 248 = 459.$$

Notice that,

$$\sum_{i=6}^{8} E_i - (\sum_{i=3}^{5} D_i + \sum_{i=0}^{2} A_i) = 459 - (88 + 12) = 495 - 100 = 359.$$

Subtracting $R^{(4)} = 20$ of Einstein's gravity tensor from the final result, one finds,

$$359 - R^{(4)} = 359 - 20 = 339,$$

Surprisingly this number is exactly equal to the number of states or dimensions of the holographic boundary of $\varepsilon^{(\infty)}$ theory, namely K($\Gamma_c(7)$). Following the preceding discussion, one can deduce that the A-D-E hierarchy, which can give interesting explanations for some features of the standard model [23, 31].

**6. A-D-E hierarchy and M-theory**

M-theory is the master non-perturbative theory that unifies the five perturbative ten-dimensional superstring theories into one theory of eleven–dimensional (D=11) supergravity. M-theory describes supersymmetric extended objects with two spatial dimensions (D=2) that are called super-membranes, and five spatial dimensions (D=5) that are called super-five-branes, which subsume all five consistent string theories. These 7 dimensions are considered to be the hidden or compactified dimensions of M-theory [1, 2].

Let us return to our sum of A-D-E hierarchy [30]. If we subtract $D^{(11)}$ super-gravity from the hierarchy sum, we find,

$$559 - D^{(11)} = 559 - 11 = 548.$$

Based on $\varepsilon^{(\infty)}$ theory [9], the number 548 can be regarded as particles like state, following the methodology of the three steps symmetry breaking, one finds,

$$(((548/2)/2)/2) = \frac{548}{8} = 68.5$$

particles. To find the integer value of N(SM), let us subtract $D^{(7)}$ from 559, namely,

$$559 - D^{(7)} = 559 - 7 = 552,$$

where, $D^{(7)} = D^{(11)} - D^{(4)}$ is the compactified section of M-theory. That way, one finds,

$$\frac{552}{8} = 69 \ particles.$$

Moving to the classical form of $E_8 \otimes E_8$ string theory, which has 496 massless state gauge boson, one finds

$$559 - |(E_8 \otimes E_8)| = 559 - 496 = 63$$

, which is equal to non-supersymmetric value deduced by EL-Naschie for the number of SM elementary particles [11]. And with agree with the classical heterotic string theory as explained by M. Green [3].

## 7. $\varepsilon^{(\infty)}$ theory, symmetry breaking and SM missing particles

Over the last two decades, string theory has been the preeminent model for physics beyond the standard model [7, 32]. Indeed, as it is known in the literature, String Theory is assumed to be the ultimate theory of everything. There are in fact several string models, the dominant model so far is that of Heterotic string theory in which $E_8$ plays an essential role. The word Heterotic means that the string theory is a subtle interweaving of the original bosonic theory of 26 dimensions and superstring theory of 10 space time dimensions [32].

Using Heterotic string theory, the 26 dimensions can be reduced to 4 dimensions in two steps. First of all, 16 of the original 26 dimensions must be compactified, and then 6 of the remaining 10 dimensions must be compactified as well in order to get down to our apparent 4-dimensional observed universe.

The heterotic string theory is described by the symmetry group $E_8 \otimes E_8$ that has 496 dimensions, with 16 commutative dimensions. The dual subspace to these 16 commutative dimensions are the 16 dimensions in which the remaining non-commutative 480 vectors carrying charges, 16 charges for each vector, which are correspond to different particle types [33].

The geometry of the 16-dimensional space time is well described by Coexter graph. This graph describes 16 basic mirrors that create the finite reflection group of $E_8 \otimes E_8$, which is usually Called Coexter reflection group. In particle physics, these reflections are responsible to force one "force particles" to another "matter particle". In other words it creates super-symmetry (SUSY) that takes one boson to another fermion and vice versa [33-35].

Now, let us look on the connection between the sphere backing and string theory, one can find the kissing number in the 10- dimensions, which is the dimensions of superstring theory to be 336. Thus from previous interpretation of $\Gamma(7)$ and $R^{(8)}$, one can give 336 particle like states a new interpretation as kissing numbers of 9-dimensional space time sphere packed around central sphere living on 10 – dimensional superstring space [36].

Following the preceding argument, one can assume that the kissing numbers of different Lie groups can be regarded as elementary particles that live on the manifolds of different lie groups [34, 36].

Now, let us have a look at the principle of symmetry breaking that was first assumed by Pierre Curie. He assumed that on the occurrence of a phenomenon in medium, the original symmetry group of the medium must be lowered than the symmetry group of the phenomenon by the action of some causes. According to this scenario, symmetry breaking is what creates the phenomenon.

The most familiar symmetry breaking is that from the string theory scale $E_8 \otimes E_8$ to the standard model scale, passing through the well-known symmetry group $E_6 \otimes E_6$ [37]. Here we use 480 massless bosons that correspond to the kissing number of Heterotic string theory $E_8 \otimes E_8$, namely $|K(E_8 \otimes E_8)| = K(E_8) + K(E_8) = 240 + 240 = 480$, and the special linear group SL(2,7) which is the symmetry group of the holographic boundary of $\varepsilon^{(\infty)}$ theory [12] to calculate the number of the elementary particles in the standard model. The dimension of the symmetry group SL(2,7) can be obtained by the following formula [30],

$$Dim\ SL(2,N) = N(N^2 -1,)$$

Setting N=7, one can find

$$Dim\ SL(2,7) = 7\ (7^2-1) = 336,$$

Which is identical to Klein's modular curve. Based on the preceding discussion, we are in position to make important symmetry breaking relation in particle physics.

$$|K(E_8 \otimes E_8)| \longrightarrow |K(E_6 \otimes E_6)|$$

That symmetry breaking relation can be obtained by subtracting $|SL(2,7)|$ from $|K(E_8 \otimes E_8)|$, consequently,

$$|K(E_8 \otimes E_8)| - |SL(2,7)| = 480 - 336 = 144.$$

Notice that,

$$|K(E_6 \otimes E_6)| = K(E_6) + K(E_6) = 72 + 72 = 144.$$

In that way, one finds

$$|K(E_8 \otimes E_8)| - |SL(2,7)| = |K(E_6 \otimes E_6)|.$$

Following the preceding plausible assumption that Coexter reflection group of $E_8 \otimes E_8$ creates super-symmetry taking one boson to fermion and vice versa, and 336 are corresponding to particle physics, one can say that previous equation represents SUSY breaking of our model. Finally we remove the left-right (L-R) symmetry which corresponds to non-super-symmetric SU(5) grand unification theory of Glashow-Georgi

[7], that way one can find the number elementary particles in an extended standard model,

$$N(SM) = \frac{|K(E_6 \otimes E_6)|}{2} = \frac{144}{2} = 72 \text{ particle.}$$

The same result was obtained by L. M. Crnjac [24], using the critical dimension of bosonic string theory $D^{(26)}$, following [10], namely

$$N(SM) = \frac{576}{8} = 72 \text{ particle.}$$

Now, let us use $|SL(2,7)|_c$ which is the $\varepsilon^{(\infty)}$ compactified version of $|SL(2,7)|$, following the same preceding scenario, one finds,

$$|K(E_8 \otimes E_8)| - |SL(2,7)|_c = 480 - 339 = 141.$$

The next step is to break the symmetry of our model to be in contact with the SM scale and that can be done by subtracting the kissing number of $E_6$, which is the symmetry group that is responsible about the final step of symmetry breaking [9,10], from 141, namely, El-Naschie's well-known results:

$$141 - K(E_6) = 141 - 72 = 69 \text{ particles.}$$

As mentioned before, sixty particles are confirmed experimentally, one massless graviton, and the remaining eight components can be regarded as a degree of freedom of the 2 complex $SU(2)_L$ Higgs doublet model which is assumed by MSSM [3]. Three of these eight components were absorbed to give the W and Z gauge boson their masses, leaving 5 degree of freedom. Two charged Higgs boson, one CP-odd neutral Higgs boson, and 2 CP-even neural Higgs boson [3].

**8. Higgs mass, kissing numbers and hierarchy symmetry breaking**

The most familiar symmetry breaking the $E_8 \otimes E_8 = 496$ symmetry of string theory to $SU(3) \times SU(2) \times U(1) = 12$ symmetry of the standard model is a series of hierarchical symmetry breaking in which $E_8 \otimes E_8$ breaks to $E_8$. Then $E_8$ breaks to $E_6$, which by turn breaks to the standard model scale [35, 38], namely

$$E_8 \otimes E_8 \longrightarrow E_8,$$

$$E_8 \longrightarrow E_6 \longrightarrow E_5 \longrightarrow E_4$$

where $E_5$ corresponds to the lie group $SO(10) = \frac{10(10-1)}{2} = 45$, and $E_4$ can be recognized as Georgi-Glashow grand unification theory $SU(5) = 5^2 - 1 = 24$ of electromagnetic weak and strong gauge forces with gauge group $SU(3)_{color} \times SU(2)_L \times U(1)_Y$ as a maximal subgroup of $SU(5)$ [7]. It is noted that, $E_7$ is skipped from the scale. This was a deliberate omission, because $E_7$ is incompatible in a certain sense, with the chiral fields. This means that the weak charge, which distinguishes between right and left – handedness, was not be properly accounted for [35]. The best way to explain this scenario is by considering the corresponding kissing number of each lie group, i.e.

$$K(E_8) \longrightarrow K(E_6) \longrightarrow K(E_5) \longrightarrow K(E_4)$$

where, $K(E_8) = 240$, $K(E_6) = 72$, $K(E_5) = 40$ and $K(E_4)=20$.

Following that scale of symmetry breaking, one notices that

$$K(E_8) - K(E_6) = 168,$$

$$K(E_6) - K(E_5) = 32, \text{ and}$$

$$K(E_5) - K(E_4) = 20.$$

In what follows we will attempt to interpret the physical meaning beyond this symmetry breaking numbers.

The basic rule of particle interactions is that fermions interact by exchanging gauge bosons. As a consequence of the self – duality of $E_8$- lattice, there is no distinction between fermions and bosons. The difference between them is observed when $E_8$ lattice is projected down to a lower dimensional lattice [34]. This projection is necessary to make contact with standard model. Here we can see that $K(E_8)$ breaks to $K(E_6)$ by subtracting 168. This number can also be regarded as elementary particles. It is exactly equal to the number of automorphism of Klein modular curve. The degree of freedom of this curve is given by twice the value of automorphism (168)(2)=336 that we regard as the dimension of the curve [36]. Thus 168 can be regarded as degree of freedom, dimension or coupling constant after one symmetry breaking 336/2=168 [38].

The Higgs mass is the only unknown parameter in the symmetry breaking sector of the standard model [1]. As $E_8$ self-dual lattice is projected down to $E_6$ lattice, the difference between bosons and fermions appears to be associated with the creation of Higgs field, and consequently Higgs field quanta, which are called Higgs boson. That way, one can say that the Higgs mass $m_H$ is created through the preceding projection, and it is numerically equal to the symmetry breaking number 168.

The electron volt units system plays a fundamental role and penetrates deep into E-infinity theory and could not be taken out of it without obscuring the theory [15]. In particular, El-Naschie showed that according to E-infinity theory there is a possibility that each dimension can correspond to a mass so it can be represented in electron volt units system, for instance we have [21]

1- The mass of the expectation $\pi$ - meson could be calculated in terms of the inverse fine structure constant $\bar{\alpha}_o$ gauged in MeV.

$$\langle \pi \rangle = \bar{\alpha}_o \text{ MeV}$$

We recall that in E-infinity theory we known that $\bar{\alpha}_o$ can be considered as a dimension [9].

2- The mass of the expectation k - meson could be calculated in terms of $E_8 \otimes E_8$, the dimension of the heterotic string theory:

$$\langle m_k \rangle = Dim\ (E_8 \otimes E_8)\ MeV$$

This symmetry between mass and dimension is considered to be a type of duality in which the mass could be represented in terms of dimension.

Following the preceding discussion, 168 can be equal to the mass of the missing Higgs boson $m_H$ interms of GeV [38].

Now, let us transfer to the second step of symmetry breaking, here $K(E_6)$ breaks to $K(E_5)$ and the number 32 appears as a result of this symmetry breaking. Needless to say again that we regard 32 as dimension or degree of freedom, in fact this number is exactly equal to the dimension of the superspace n=(8)(4) = 32 which is a sub-space of $\varepsilon^{(\infty)}$ theory space [12]. Removing such number that equals to the dimension of spacetime in which supersymmetry works, makes no doubt for us that this step includes supersymmetry breaking.

Moving to the final step of that scheme we come into contact with the standard model. Thus, far we had only 3 fundamental forces namely electromagnetic, weak and strong forces as mentioned before in this paper. Therefore, the next step must involve the $R^{(4)} = 20$ of Einstein's gravity tensor. This is obvious throughout the process of $K(E_5)$ breaking to $K(E_4)$. Here the number is 20 which is exactly equal to the number of independent components of Riemannian tensor in 4 dimensions [13],

$$R^{(n)} = \frac{n^2(n^2-1)}{12},$$

setting n=4 and one finds that

$$R^{(4)} = \frac{4^2(4^2-1)}{12} = 20$$

which represents the gravitational force field.

Following the preceding symmetry breaking scenario, one can assume the following hierarchy [41- 43],

$$|D(E_8)| = 248, \quad |D(E_6)| = 78, \quad |D(E_5)| = 45, \quad and \quad |D(E_4)| = 24$$

Now, let us consider the sum,

$$|D(E_8)| + |D(E_6)| + |D(E_5)| + |D(E_4)| = 248 + 78 + 45 + 24 = 395.$$

This sum of the symmetry breaking hierarchy components can be regarded as dimension or degree of freedom. This degree of freedom is considered to describe the physics of gauge boson from the high energy scale to the SM scale. In other words one can say that this number is related to the whole mass of the gauge boson sector of the SM. According to the previous discussion and the mass dimension duality generated by $\varepsilon^{(\infty)}$ theory [15, 16, and 38], this number can be expressed in eV units to be corresponding to gauge bosons' masses, consequently we find the following equation,

$$\sum (symmtry\ breaking\ components) = \sum (gauge\ bosons\ masses)$$

Therefore, one finds that,

$$\sum (symmtry\ breaking\ components) = \sum_{spin=0} (m) + \sum_{spin=1} (m)$$

The first term in the right hand side of the equation represents spin zero Higgs boson. The second term represents the contribution of all spin 1 gauge boson, and this term can be reduced to approximately the sum of $W^{\pm}$ and $Z^0$ masses [15].

Thus, we can write,

$$395 = m_H + m_Z + 2(m_W)$$

Setting,

$$m_Z = 91 \, GeV \quad, and \quad m_W = 80 \, GeV,$$

In the above equation, one finds,

$$m_H = 395 - 91 - 2(80) = 144 \, GeV,$$

Which is the exact new upper limit for the mass of the Higgs boson considered by the Fermi lab Tevatron collider based on the new values for the W boson mass and the top quark mass with 95 percent probability [44].

As mentioned before,

$$|K(E_6 \otimes E_6)| = 144,$$

which is numerically equal to the value of the Higgs boson mass, by this way one can say,

$$m_H = |K(E_6 \otimes E_6)| = 144 \text{ GeV}.$$

Based on the pervious calculations of the Higgs boson mass, one can calculate the arithmetic mean of both 168 Gev and 144 GeV, as follows,

$$\langle m_H \rangle = \frac{168 + 144}{2} = 156 \, GeV.$$

We note that,

$$|D(E_6 \otimes E_6)| = 156,$$

Exactly as mentioned before, and one can write that,

$$\langle m_H \rangle = |D(E_6 \otimes E_6)| = 156\, GeV.$$

The conformation of the final results of the preceding equations may leads us to consider special peculiarities of the lie groups $E_6 \otimes E_6$ and $E_6$ which play a vital role in the symmetry breaking process. In fact EL-Naschie considers the super imposition of all exceptional lie groups [43].

## 9. Group theory and fine structure constants

### 9.1 Estimation of the inverse quantum gravity coupling constant $\bar{\alpha}_g$

Following $\varepsilon^{(\infty)}$ theory, El-Naschie calculated the inverse quantum gravity coupling constant $\bar{\alpha}_g$ from [45]

$$\bar{\alpha}_g = \frac{[DimE_8 \otimes E_8] - k^2}{[Dim(SU(3) \otimes SU(2) \otimes U(1)] - \phi^4} = \frac{(469 - k^2)}{\sqrt{\bar{\alpha}_o}} = 42 + k = 42.36067977.$$

As mentioned before, $\Gamma_c$ (7) is the holographic boundary of $\varepsilon^{(\infty)}$ theory. Thus instead of using 496 massless boson we will consider only 336 higher dimensional Gluon –quark- like states [45]. But, the SM has only Dim SU(3) =9-1 = 8 Gluons. Thus, we can estimate the $\bar{\alpha}_g$- value from

$$\bar{\alpha}_g \approx \frac{(336)}{8} \approx 42$$

which is close to the exact value, to obtain the exact value we have to include the $\varepsilon^{(\infty)}$ theory transfinite corrections to find that,

$$\bar{\alpha}_g = \frac{Dim\Gamma_c(7)}{DimSU(3)} = 42.36067977.$$

Note that 42+2k may be regarded as the number of elementary particles in the standard model [45]. While 26+k are gauge bosons. The total sum is nearly 69 as it should be following EL-Naschie's theory [25]

In what follows, we will use the previous interpretation of the kissing number as elementary particles together with the exceptional lie groups $E_8$, $E_7$ and $E_6$ to estimate the numerical value of $\bar{\alpha}_g$ from the relation,

$$K(E_8) - K(E_7) - K(E_6) = 240 - 126 - 72 = 42.$$

Now, to obtain the exact value of $\bar{\alpha}_g$, we have to consider the exact transfinite values of $K(E_8)$ and $K(E_6)$, that can be done by adding the dimension value of each lie group multiplied by the transfinite correction term k, where $k = \phi^3(1-\phi^3) = 0.18033989$ and $\phi = 0.618033989$. Thus one finds,

$$K_c(E_8) = 240 + 8k \quad \text{and} \quad K_c(E_6) = 72 + 6k$$

Consequently, the exact value of $\bar{\alpha}_g$ is given by

$$(240 + 8k) - 126 - (72 + 6k) = 42 + 2k.$$

Exactly as it should be.

## 9.2 Estimation of the inverse of electromagnetic fine structure constant $\bar{\alpha}_o$

Electromagnetic fine structure constant has been always a mystery ever since it was discovered. It is a magic number that plays a very important role in many fields related to the human environment, such as fiber optical light communication channels and digital photo devices [9]. The usual definition of the fine structure constant can be written as

$$\alpha = \frac{e^2}{4\pi\varepsilon_o c\hbar} = 7.2973525568(24) = \frac{1}{137.03599911(46)}.$$

where, e is the elementary charge $\hbar$ is the reduced plank's constant that equal to $\frac{h}{2\pi}$, c is the speed of light, and $\varepsilon_o$ is the permittivity of the free space.

Following the framework of E-infinity theory, using the transfinite exact value of symmetry group $|E_8 \otimes E_8|_c = 496 - k^2 \approx 496$ of string theory and exact value of special linear group $|SL(2,7)|_c = 336 + 16k = 338.8854382 \approx 339$. El-Naschie calculated, for the first time, the exact theoretical value for the inverse of electromagnetic fine structure constant $\bar{\alpha}_o$ [46].

$$\bar{\alpha}_o = |E_8 \otimes E_8|_c - |SL(2,7)|_c - R^{(4)} = (469 - k^2) - (336 + 16k) - 20 = 137.0820393$$

where $R^{(4)}$ is the 4-dimensional tensor.

Next, we introduce a new but simple way to estimate the numerical value for the inverse of electromagnetic fine structure constant $\bar{\alpha}_o$. The main fact used in the present estimation that the kissing number of Heterotic string theory $|K(E_8 \otimes E_8)| = 480$ can be regarded as a degree of freedom in exactly the same way the number of elementary particles on the standard model N(SM)= 68.54102 ≈69 can be regarded also as quasi degree of freedom. Start by adding all the degrees of freedom, thus the sum is

$$|K(E_8 \otimes E_8)| + N(SM) = 480 + 68.54102 = 548.54102.$$

Following El-Naschie [13], one finds,

$$\bar{\alpha}_o = \frac{548.54102}{(D^{(4)} = 4)} = 137.135225$$

which is quite close to the value obtained earlier also by El Naschie [46].

## 10. Summary and Conclusion

In the present work we have established a new symmetry breaking relation using E-infinity theory. This relation is

$$|K(E_8 \otimes E_8)| - |SL(2,7)| = |K(E_6 \otimes E_6)|.$$

Assuming (L-R) symmetry breaking, one finds the number of elementary particles in the extended standard model to be

$$N(SM) = \frac{|K(E_6 \otimes E_6)|}{2} = \frac{144}{2} = 72.$$

In addition we have discussed the hierarchy symmetry breaking scenario using the kissing numbers of the exceptional lie groups, namely

$$K(E_8) \longrightarrow K(E_6) \longrightarrow K(E_5) \longrightarrow K(E_4)$$

We have estimated within the framework of $\varepsilon^{(\infty)}$ theory the transfinite corrected value of the inverse quantum gravity coupling constant $\bar{\alpha}_g$, using the equation

$$K_c(E_8) - K(E_7) - K_c(E_6) = (240 + 8k) - 126 - (72 + 6k) = 42 + 2k.$$

Also, the numerical value of the inverse of electromagnetic fine structure constant $\bar{\alpha}_o$ has been estimated in a simple way, namely

$$\bar{\alpha}_O = \frac{K(E_8 \otimes E_8) + N(SM)}{(D^{(4)} = 4)} = 137.135225.$$

We thus conclude that at least one more particle probably the Higgs must be found experimentally in the near future, making the number of particles in the standard model 61. However the most likely number to be discovered is 6 more elementary particles, namely 66 all in all.

The maximum number at energy below one Tesla is found by El-Naschie, namely 69 particles, i. e 9 more particles must be found [25].

Nonetheless at higher energy involving super-symmetric parameter, the value 72 predicted by Marek- Crnjac and El-Naschie is possible. In fact at ever higher values there are possibilities with the $E_8$ to E-infinity scenarios [29] that a total of 80 or 84 elementary particles exist. This means we could discover as many as 20 or even 24 more particles. Clearly Experimental High energy physics still a long way to go [13]

## 11. References

[1] D.H. Perkins. Introduction to high energy physics. Cambridge ,2000.

[2] B.R. Martin. Shaw, G., Particle physics. Manchester, 1997.

[3] S. Dawson. The MSSM and way it works. hep- ph/9712464v1.